\documentclass[pss,fleqn,embeddedheads]{w-art}
\usepackage{times}
\usepackage[]{graphicx}
\usepackage{subfigure}

\begin{document}
\renewcommand{\copyrightyear}{2005}% uncomment to change the copyrightyear.


\title[Onset of metallic ferromagnetism in a doped spin-orbital chain]{Onset
  of metallic ferromagnetism in a doped spin-orbital chain}

\author[Daghofer]{Maria Daghofer \inst{1,}\footnote{Corresponding
     author: e-mail: {\sf daghofer@itp.tu-graz.ac.at} }} 
\address[\inst{1}]{Institute for Theoretical and Computational Physics, 
                   TU Graz, Patersgasse 16, A-8010 Graz, Austria}
\author[Ole\'{s}]{Andrzej M. Ole\'{s} \inst{2,3}}
\address[\inst{2}]{M. Smoluchowski Institute of Physics, Jagellonian 
                   University, Reymonta 4, PL-30059 Krak\'ow, Poland}
\address[\inst{3}]{Max-Planck-Institut f\"ur Festk\"orperforschung,
                   Heisenbergstrasse 1, D-70569 Stuttgart, Germany}
\author[von der Linden]{Wolfgang von der Linden \inst{1}}
\begin{abstract}
Starting from a spin-orbital model for doped manganites, we investigate
a competition between ferromagnetic and antiferromagnetic order  
in a one-dimensional model at finite temperature. The magnetic and 
orbital order at half filling support each other and depend 
on a small antiferromagnetic superexchange between $t_{2g}$ spins and 
on an alternating Jahn-Teller potential. The crossover to a metallic
ferromagnetic phase found at finite doping is partly suppressed by the 
Jahn-Teller potential which may localize $e_g$ electrons.
[published in phys. stat. sol. {\bf 242}, 311 (2005)]
\end{abstract}

\maketitle

\section{Introduction}

The transition from an insulating to metallic behavior in doped
manganites, observed as a function of temperature or filling, is one of 
the outstanding problems in the field of strongly correlated systems. 
The theoretical challenge is to understand a rich variety of metallic,
insulating, magnetically ordered and orbital ordered phases of doped 
manganites in terms of the dynamics of correlated $e_g$ electrons which 
involves their orbital degrees of freedom \cite{Dag02}. Magnetic 
interactions: spin superexchange (SE) and double exchange (DE), as well 
as orbital SE, frustrate each other and their competition leads to 
several magnetic phases, including phase separation \cite{Ali01}.  

While orbital order supports $A$-type antiferromagnetic (AF) phase in 
LaMnO$_3$ \cite{feiner99}, one expects that $e_g$ orbitals play an 
important role in the insulating ferromagnetic (FM) phase  at finite 
doping, below the crossover to a metallic FM phase \cite{Dag02}.
Here we investigate a one-dimensional (1D) model derived from the 
realistic spin-orbital model for manganites \cite{feiner99}, showing 
an interplay between spin and orbital order.

\section{ Effective orbital $t$-$J$ model at finite temperature }

We investigate the 1D orbital $t$-$J$ model for manganites for a given
configuration of core spins ${\cal S}$, 
\begin{equation}
{\cal H}({\cal S})=H_t+H_J+H_{J'}+H_\textnormal{JT},
\label{HtJ}
\end{equation}
which consists of the hopping term $H_t$, the orbital superexchange $H_J$ 
for the $e_g$ electrons, superexchange $H_{J'}$ for the core spins formed 
by $t_{2g}$ electrons, and an alternating Jahn-Teller (JT) potential 
$H_{JT}$.  

As we consider 1D chains in $z$-direction, only electrons within 
$|z\rangle\equiv\frac{1}{\sqrt{6}}|3z^2-r^2\rangle$ orbitals are mobile, 
while those within $|x\rangle\equiv \frac{1}{\sqrt{2}}|x^2-y^2\rangle$ 
orbitals are localized:
\begin{equation}
H_t=-\sum_{i}t_{i,i+1}
    ({\tilde c}_{iz}^{\dagger}{\tilde c}_{i+1,z}^{}
    +{\tilde c}_{i+1,z}^{\dagger}{\tilde c}_{iz}^{}),
\label{Ht}
\end{equation}
where $\tilde c_{iz}$ and $\tilde c_{iz}^\dagger$
are the annihilation and creation operators restricted to the Hilbert 
space without double occupancies. Due to strong Hund's coupling between  
$e_g$ and $t_{2g}$ electrons, the hopping amplitude, 
$t_{i,i+1} = tu_{i,i+1} = t \cos(\theta_{i,i+1}/2)e^{i\chi_{i,i+1}}$,
depends of the relative angle $\theta_{i,i+1}$ of the two classical core 
spins at sites $i$ and $i+1$, see e.g. Ref. 
\cite{dagotto98:_ferrom_kondo_model_mangan}. The complex phase 
$\chi_{i,i+1}$ can be neglected for a 1D chain \cite{KollerPruell2002a}. 

It is convenient to introduce orbital pseudospin operators 
$T_{i}^z=\frac{1}{2}\sigma_{i}^z=\frac{1}{2}(n_{ix}-n_{iz})$, with 
eigenstates $|z\rangle$ and $|x\rangle$. With the electron number 
operator $\tilde n_i$ restricted to the Hilbert space without double 
occupancies, the superexchange term which follows from Refs. 
\cite{feiner99} and \cite{Ole02} becomes
\begin{equation}\begin{split}\label{HJ}
H_J&= \frac{1}{5}J\sum_{i}\big(2|u|_{i,i+1}^2+3\big)
\Big( 2T_{i}^zT_{i+1}^z
-\frac{1}{2}{\tilde n}_i{\tilde n}_{i+1} \Big)
-\frac{9}{10}J\sum_{i}
\big(1-|u|_{i,i+1}^2\big){\tilde n}_{iz}{\tilde n}_{i+1,z}\\
&-J\sum_{i}\big(1-|u|_{i,i+1}^2\big)
\Big[{\tilde n}_{iz}(1\!-\!{\tilde n}_{i+1})
+(1\!-\!{\tilde n}_i){\tilde n}_{i+1,z}\Big].              
\end{split}\end{equation}
Like the kinetic energy, it depends on the direction of core spins via 
the factors $|u|^2_{i,i+1}$. The first term $\propto (2|u|_{i,i+1}^2+3)$ 
favors FM spin order for ion pairs Mn$^{3+}$--Mn$^{3+}$, while the second 
and third ones $\propto (1-|u|_{i,i+1}^2)$ favor AF order for 
Mn$^{3+}$--Mn$^{3+}$ pairs [$\propto {\tilde n}_{iz}{\tilde n}_{i+1,z}$], 
and for Mn$^{3+}$--Mn$^{4+}$ pairs 
[$\propto {\tilde n}_{iz}(1\!-\!{\tilde n}_{i+1})$]. The SE 
parameter $J=t^2/\varepsilon(^6\!A_1)$ is given by the high-spin 
excitation energy $\varepsilon(^6\!A_1)$ and the prefactors of the above
terms were chosen to preserve the relative magnitude of the different
excitations for realistic parameter values \cite{feiner99}.  Throughout 
this work, we use $J=0.125t$ and choose $t$ as unit of energy.

The third term in Eq. (1) is an AF Heisenberg interaction between the
core $t_{2g}$ spins, 
\begin{equation}
\label{HJ'}
  H_{J'} = J'\sum_{i}\big({\vec S}_i\cdot{\vec S}_{i+1}-S^2\big).
\end{equation}
As we used classical core spins ${\vec S}_i$ of unit length, their 
physical value $S=3/2$ was compensated by a proper increase of $J'$. 
Finally, we include an alternating JT potential of the form
\begin{equation}
\label{HJT}
H_\textnormal{JT}=2E_\textnormal{JT}\sum_{i}\exp(i\pi R_i)\,T_{i}^z,
\end{equation}

We employ a Markov chain Monte Carlo (MCMC) algorithm for the classical
core spins (see e.g. Ref. \cite{dagotto98:_ferrom_kondo_model_mangan}) 
combined with exact diagonalization for the electron degrees of
freedom \cite{KollerPruell2002b}. In this method, the many-particle 
Hamiltonian determined by $\{u_{i,i+1}\}$, which are in turn given 
by the core spin configuration ${\cal S}$, is solved by exact 
diagonalization taking into account its block-diagonal structure. 
The lowest states of each block are used to evaluate the trace over 
the fermionic degrees of freedom, 
$\textrm{Tr}_c\,\textrm{e}^{-\beta {\cal H(S)}}=:w({\cal S})$, which 
gives the statistical weight for ${\cal S}$ and which is sampled by the
MCMC. Autocorrelation analysis was used in order to ensure that enough
configurations were skipped between measurements.

\section{ Numerical results and discussion }

\begin{figure}[htb]
  \centering
  \hspace*{-0.4em}
  \subfigure{\includegraphics[height=0.26\textwidth]{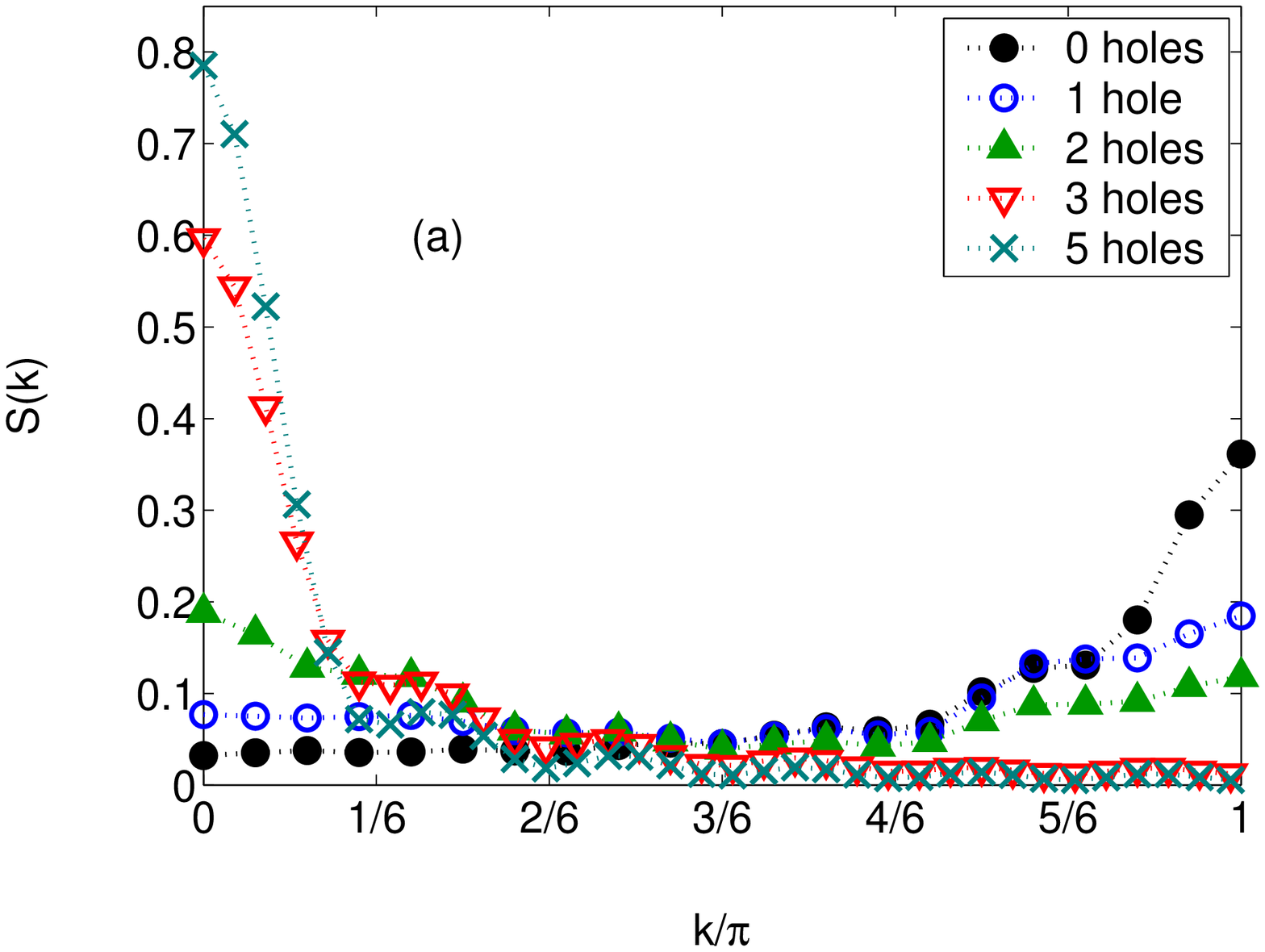}\label{ssk_Jse0.02_EJT0}}\hspace{-0.5em}
  \subfigure{\includegraphics[height=0.265\textwidth]{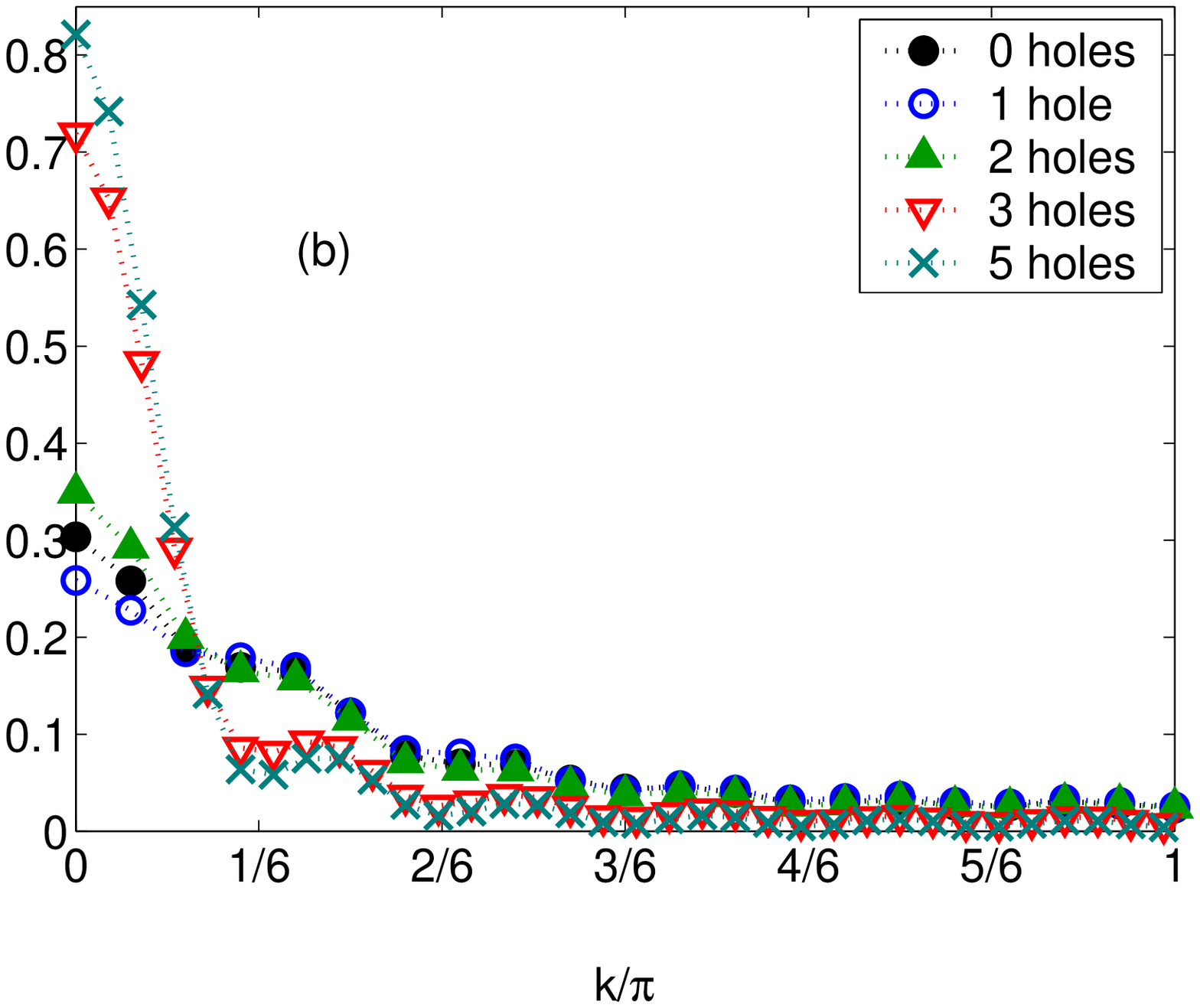}\label{ssk_Jse0_EJT0}}\hspace{-0.5em}
  \subfigure{\includegraphics[height=0.26\textwidth]{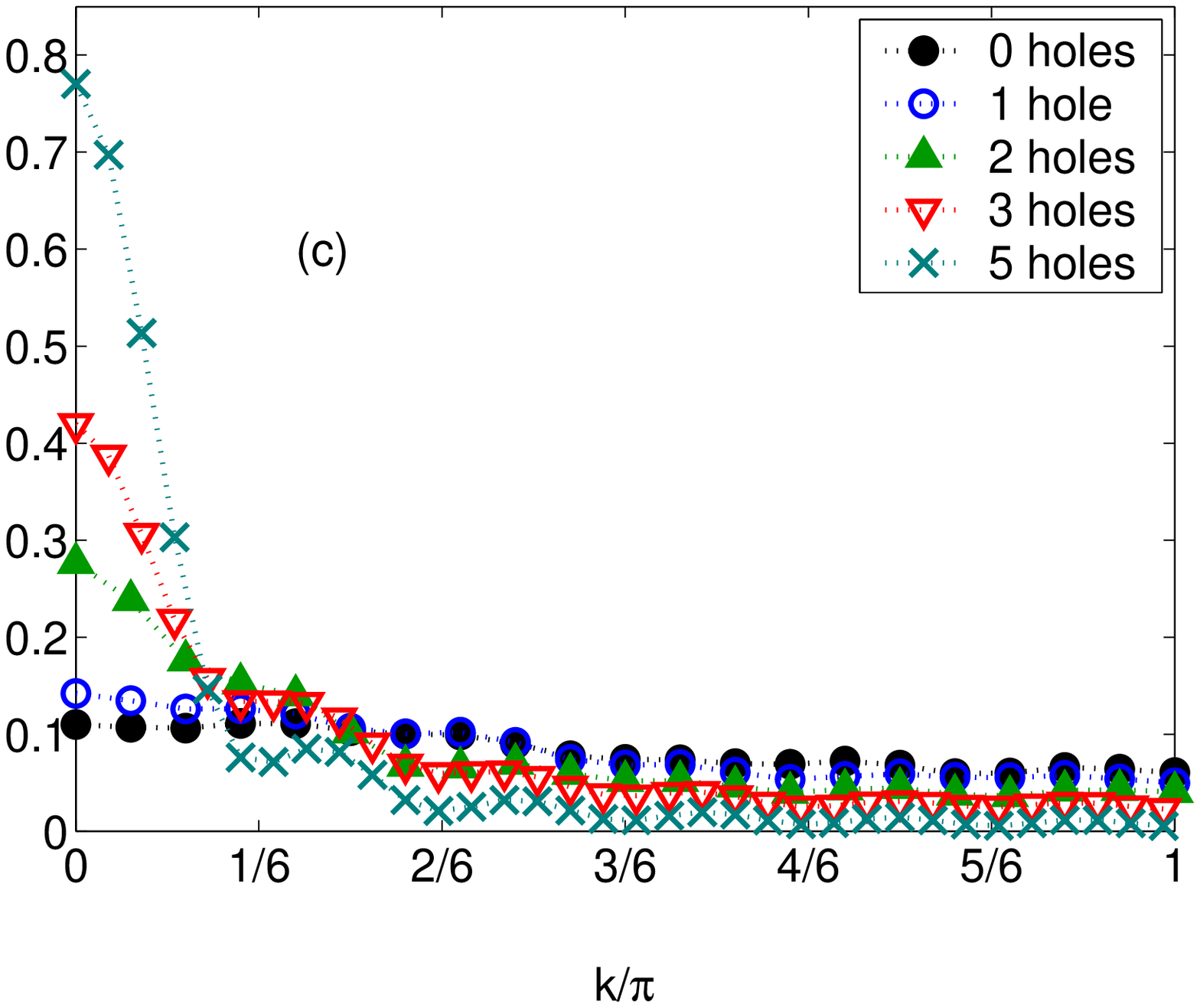}\label{ssk_Jse0.02_EJT0.1}}
  \caption{
  Core spin structure factor $S(k)$ for: (a) $J'=0.02t, E_{JT}=0$; 
  (b) $J'=0, E_{JT}=0$; (c) $J'=0.02t, E_{JT}=0.1t$, and several doping 
  levels, as obtained for an $N=12$ site chain. Parameters: $J=0.125t$
  and $\beta t=100$.
  \label{ssk}}
\end{figure}

\begin{figure}[htb]
  \centering
  \hspace*{-0.4em}
  \subfigure{\includegraphics[height=0.26\textwidth]{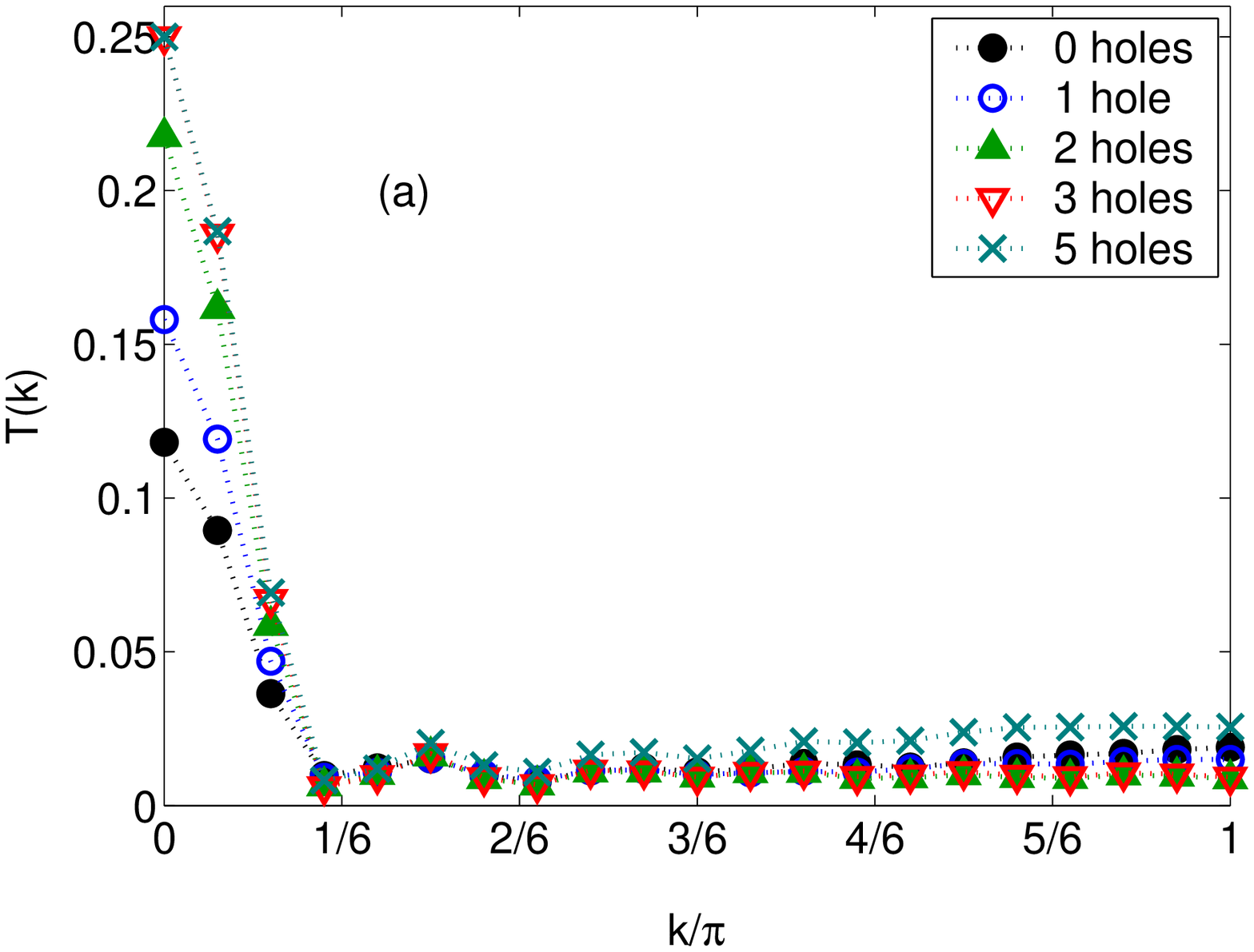}\label{ttk_Jse0.02_EJT0}}\hspace{-0.5em}
  \subfigure{\includegraphics[height=0.26\textwidth]{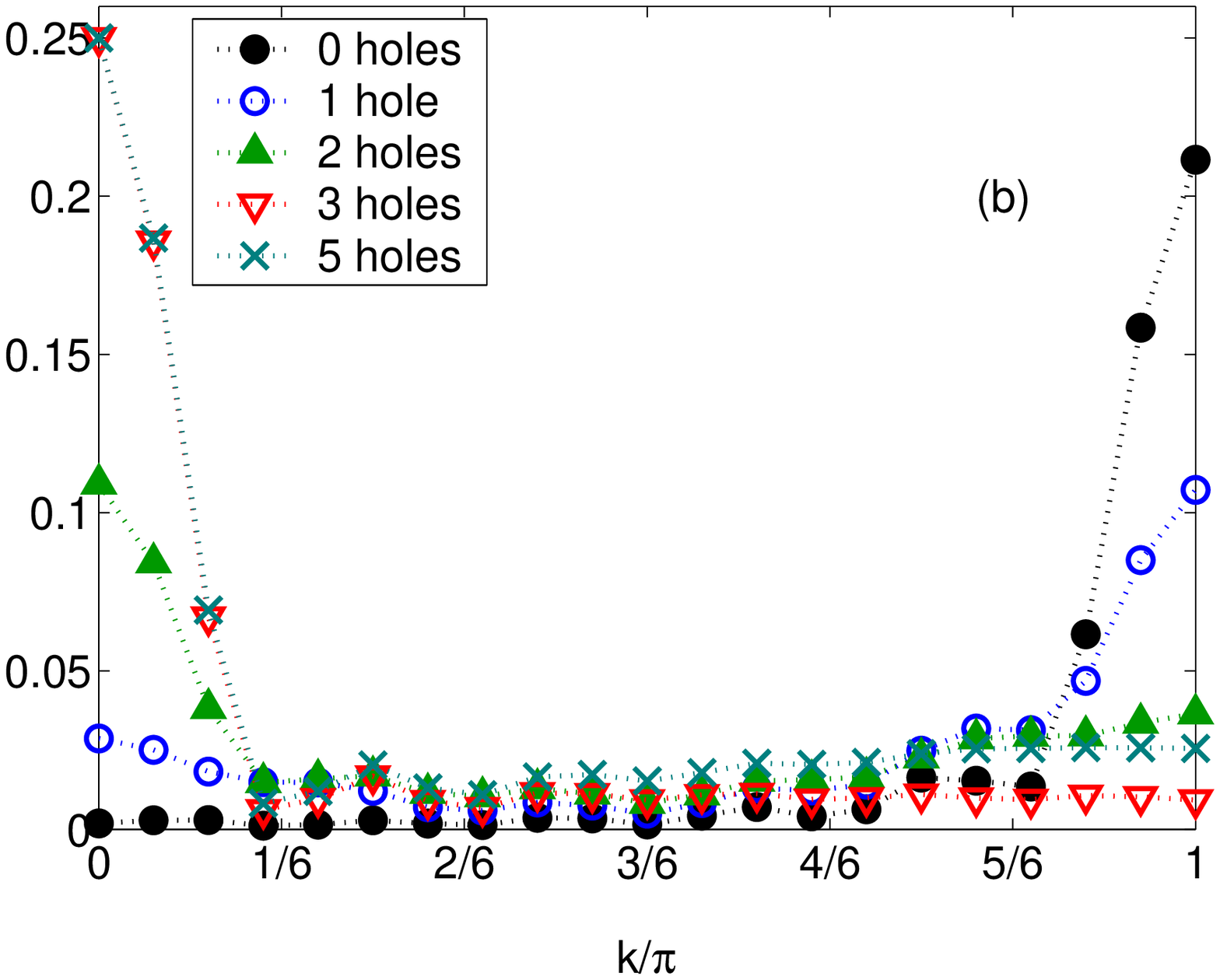}\label{ttk_Jse0_EJT0}}\hspace{-0.5em}
  \subfigure{\includegraphics[height=0.26\textwidth]{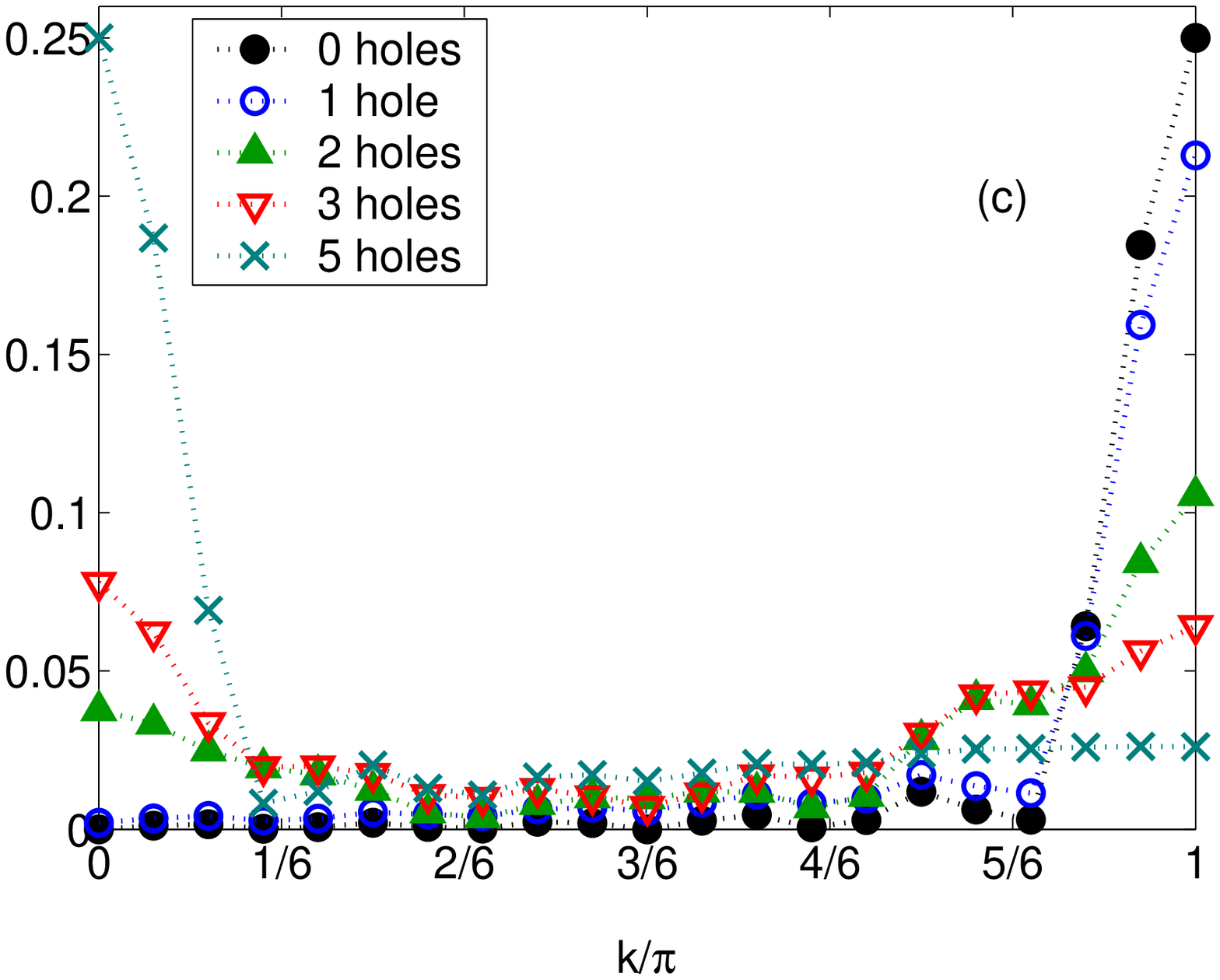}\label{ttk_Jse0.02_EJT0.1}}
  \caption{
  Orbital structure factor $T(k)$ for: (a) $J'=0.02t$, $E_{JT}=0$;
  (b) $J'=0$, $E_{JT}=0$; (c) $J'=0.02t$, $E_{JT}=0.1t$, and several 
  doping levels, as obtained for an $N=12$ site chain. 
  Parameters: $J=0.125t$ and $\beta t=100$.
  \label{ttk}}
 \end{figure}

We begin with analyzing spin and orbital correlations at increasing 
doping. Figure \ref{ssk} shows the core spin structure factor,
$S(k) = \sum_{i,j} S_iS_j\textrm{e}^{-ik(i-j)}$, for an $N=12$ chain 
and doping $0\leq x\leq 5/12$ ($x=1-n$, where $n$ is electron filling) 
near the system filled by $n=1$ electron per site (as in LaMnO$_3$), 
and for three different parameter sets at low temperature. The orbital
structure factor,
$T(k) = \sum_{i,j} T^z_iT^z_j\textrm{e}^{-ik(i-j)}/n^2$, which is 
renormalized by the squared electron density for better comparability 
of results obtained at different fillings, for the same parameters is 
shown in Fig. \ref{ttk}. Altogether, the orbital and magnetic order 
obtained at $x=0$ and $E_{\rm JT}=0$ {\it complement each other\/} --- 
polarized (ferro) $|z\rangle$ orbitals at $J'=0.02t$ induce AF spin order 
[Figs. \ref{ssk}(a) and \ref{ttk}(a)], but if the occupied $e_g$ orbitals 
alternate, FM spin order is supported instead [Figs. \ref{ssk}(b) and 
\ref{ttk}(b)]. Increasing doping induces a gradual crossover towards FM 
order due to the DE mechanism, being the most efficient in the metallic 
phase when only $|z\rangle$ orbitals are occupied. This insulator-metal 
transition is hindered by the JT potential which favors orbital 
alternation [Fig. \ref{ttk}(c)] and suppresses FM correlations
[Fig. \ref{ssk}(c)].  

\begin{figure}[htb]
   \centering
  \subfigure{\includegraphics[width=0.325\textwidth]{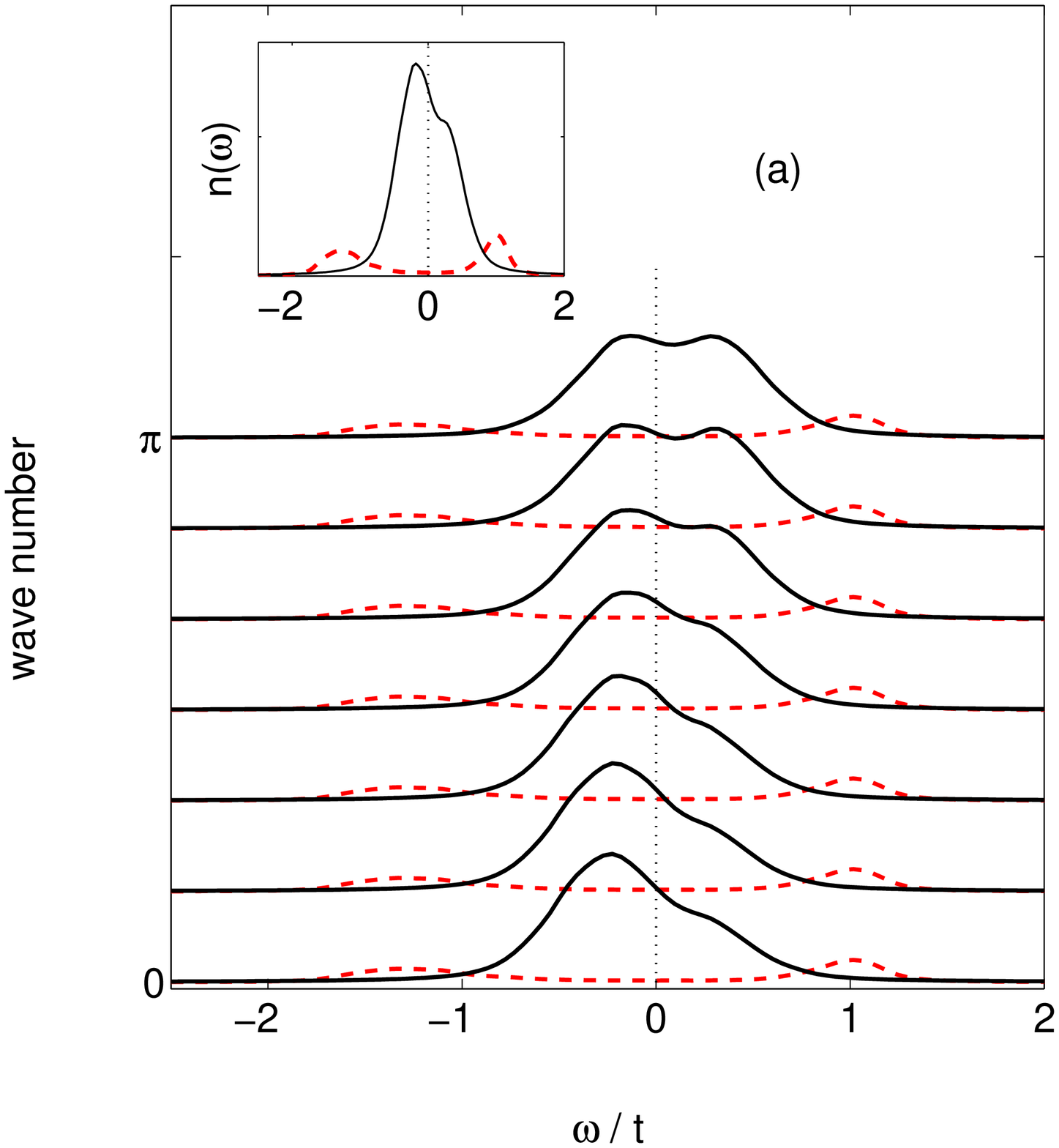}\label{Ak_b100_Jse0.02_0h}}
  \subfigure{\includegraphics[width=0.325\textwidth]{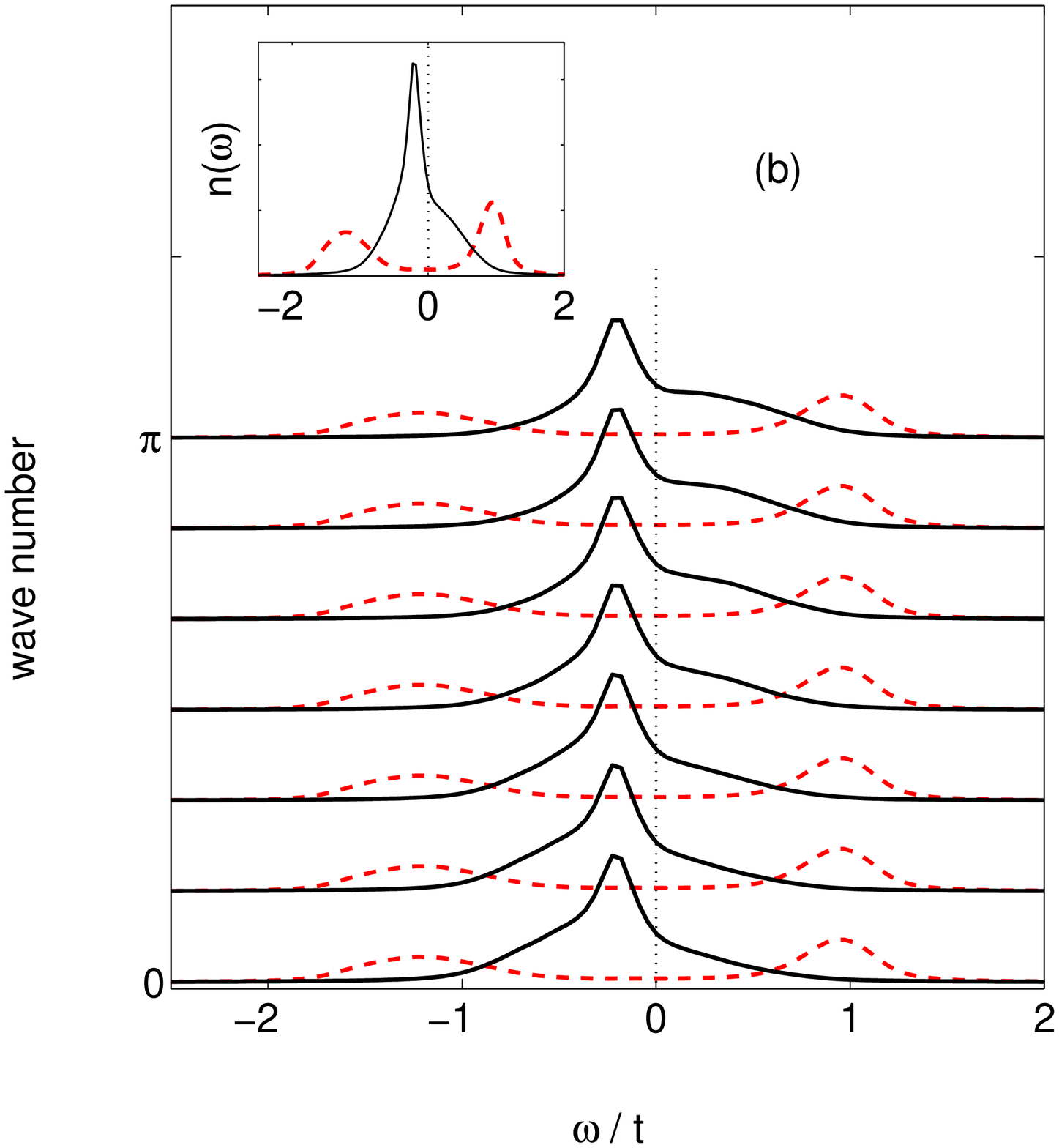}\label{Ak_b30_Jse0.02_0h}}
  \subfigure{\includegraphics[width=0.325\textwidth]{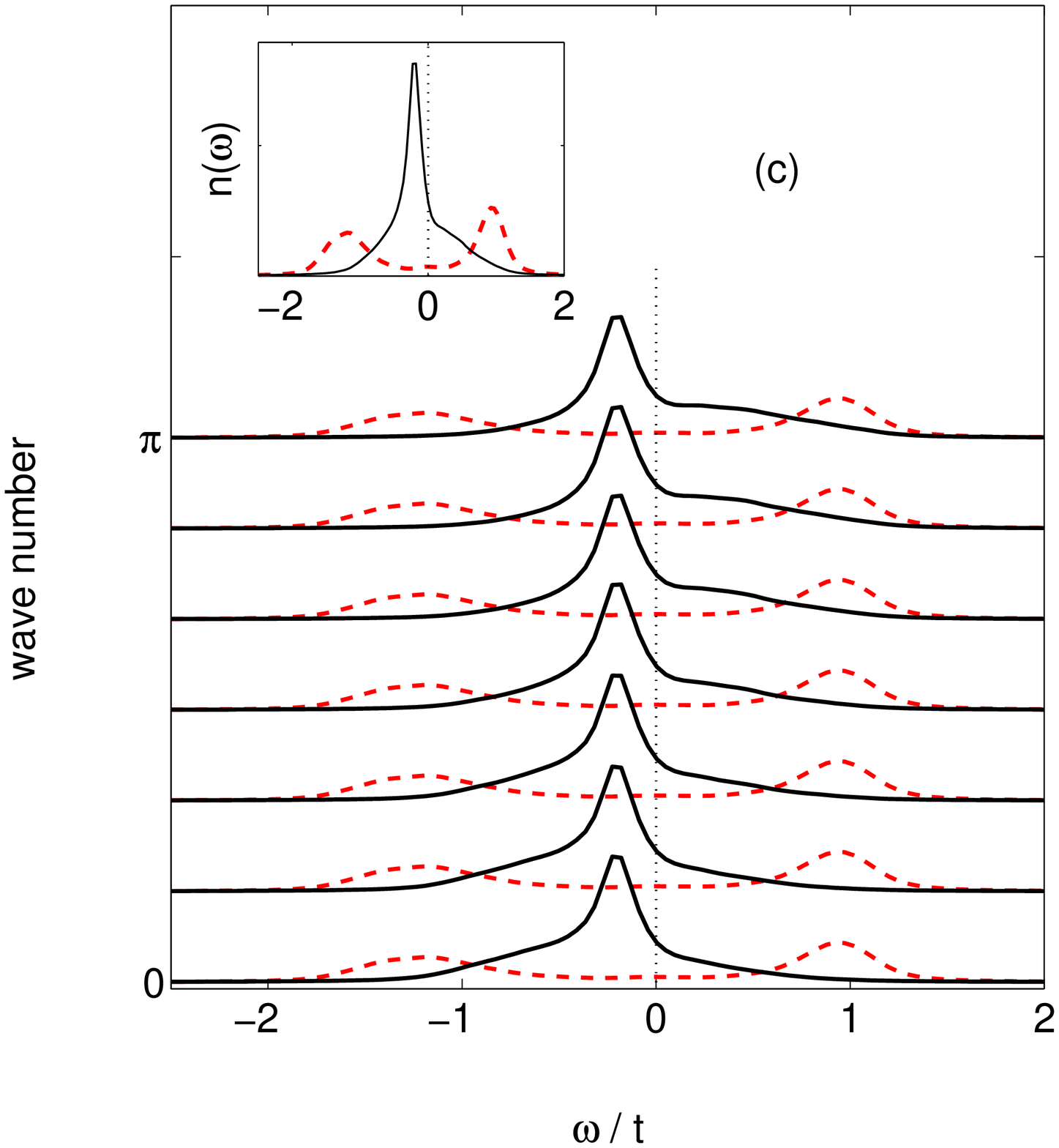}\label{Ak_b10_Jse0.02_0h}}\\[-2em]
  \subfigure{\includegraphics[width=0.325\textwidth]{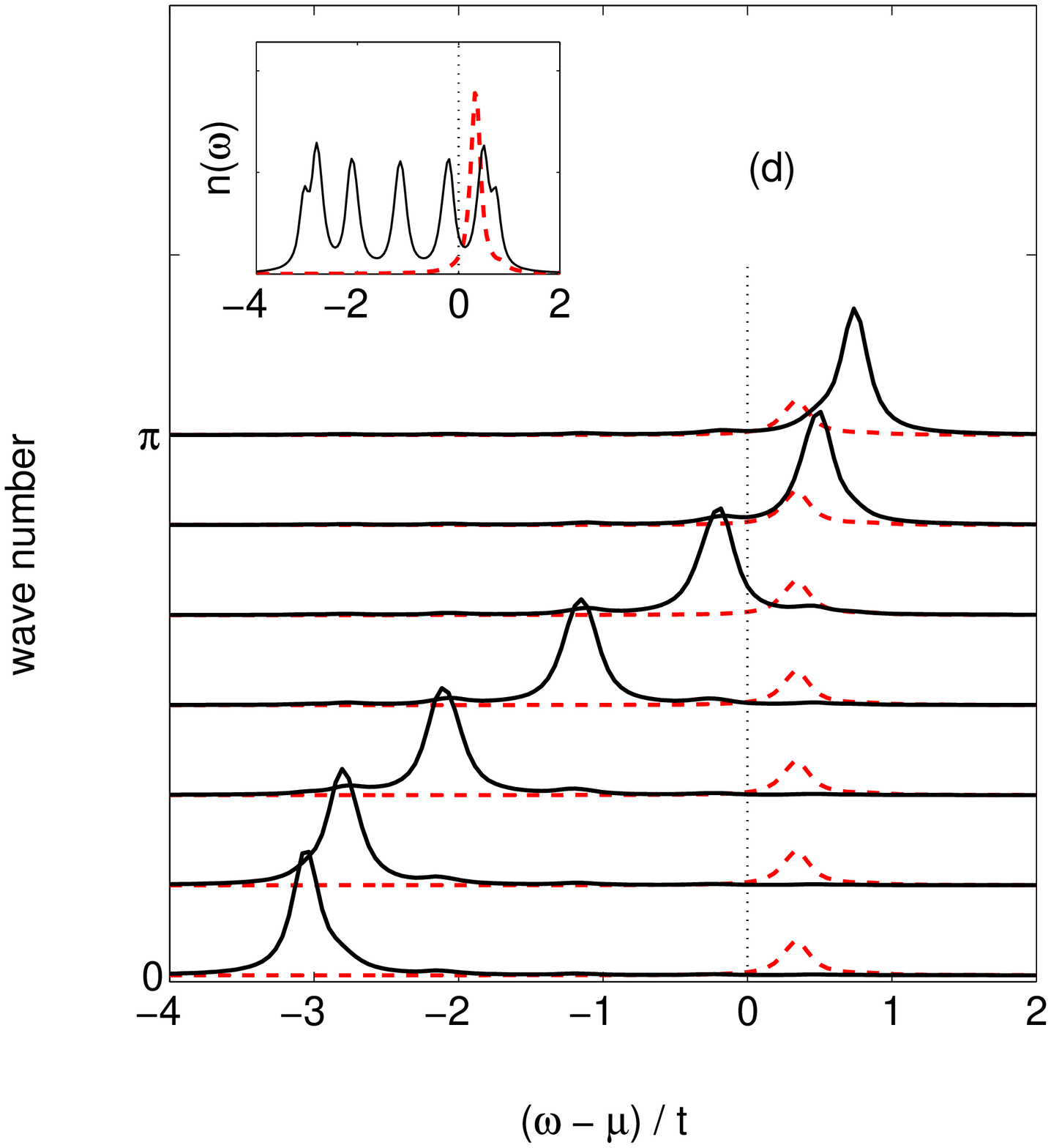}\label{Ak_b100_Jse0.02_3h}}
  \subfigure{\includegraphics[width=0.325\textwidth]{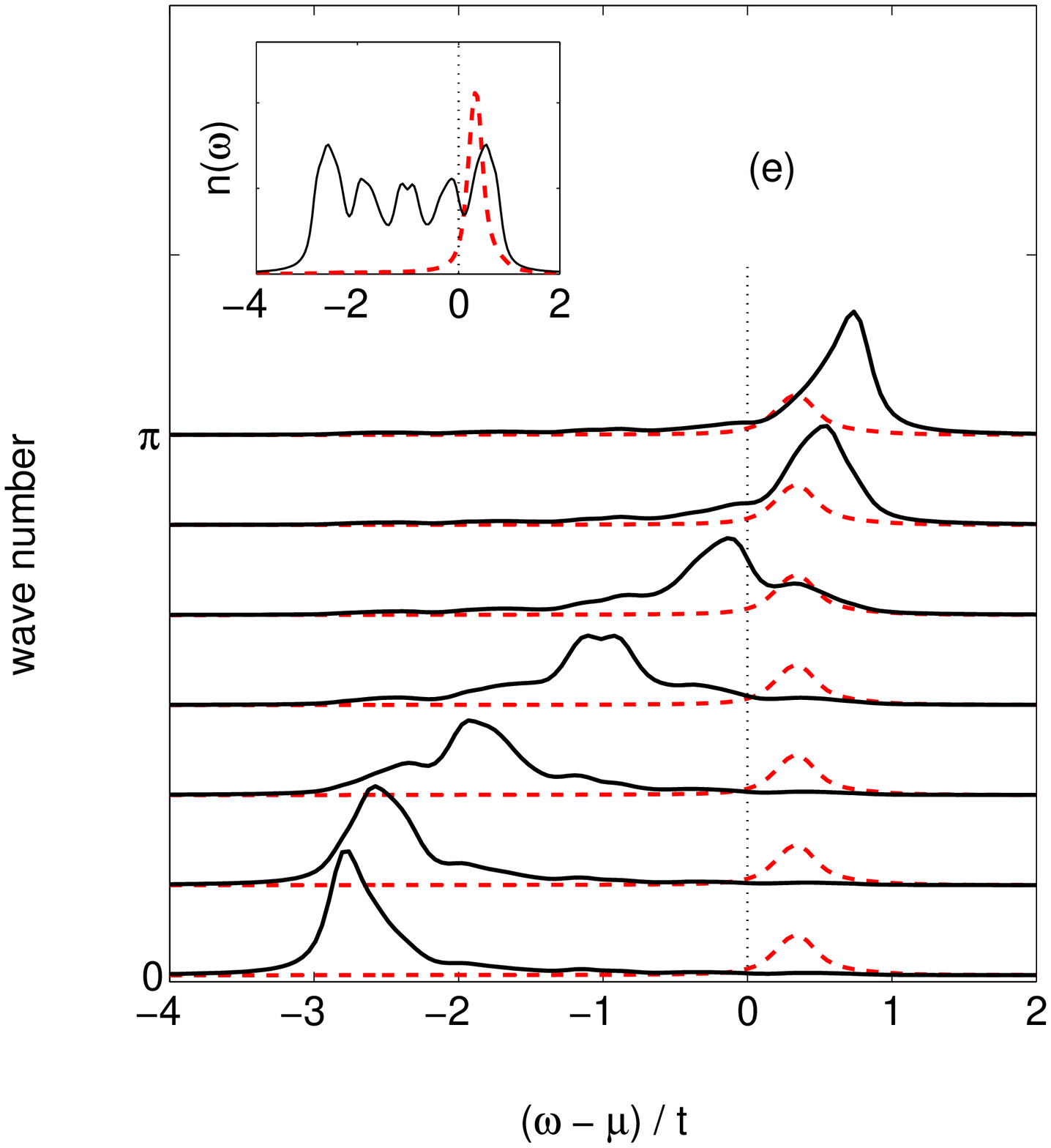}\label{Ak_b30_Jse0.02_3h}}  
  \subfigure{\includegraphics[width=0.325\textwidth]{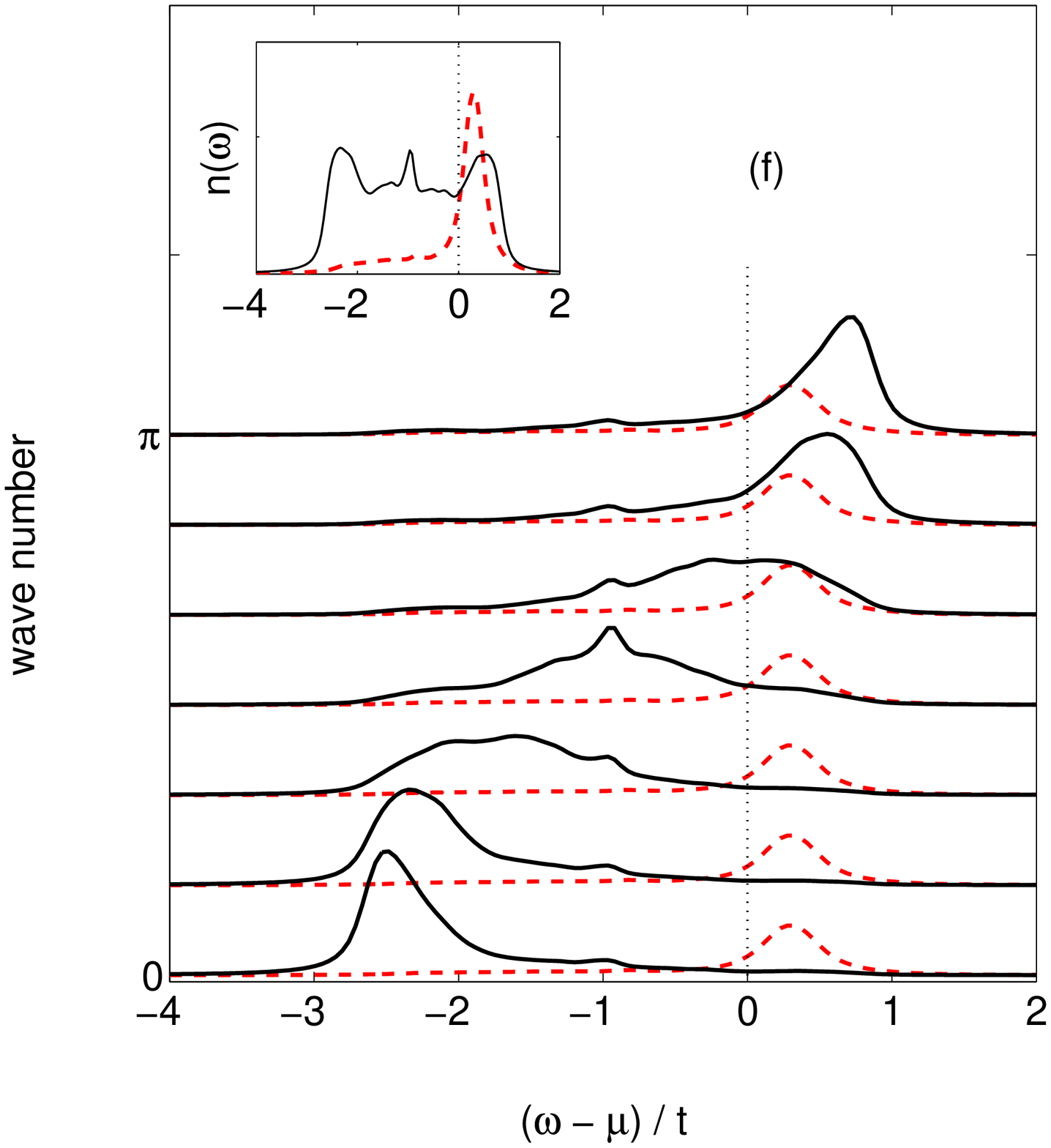}\label{Ak_b10_Jse0.02_3h}}
  \caption{
  Spectral functions $A(k,\omega)$ for an $N=12$ orbital chain (1). 
  First row --- undoped chain ($x=0$) at temperatures: (a) $\beta t=100$, 
  (b) $\beta t=30$, and (c) $\beta t=10$; 
  second row --- doped chain with three holes ($x=0.25$) at temperatures: 
  (d) $\beta t=100$, (e) $\beta t=30$, and (f) $\beta t=10$. 
  Solid and dashed lines for $|z\rangle$ and $|x\rangle$ excitations. 
  Insets show the density of states $n(\omega)$.
  Parameters: $J=0.125t$, $E_{\rm JT}=0$, $J'=0.02t$.\label{Ak_Jse0.02}
  }
\end{figure}

For a finite SE between core spins $J'=0.02t$ and in absence 
of JT potential ($E_{JT}=0$), a transition from AF to FM spin order is 
induced by doping [Fig.~\ref{ssk_Jse0.02_EJT0}], and predominantly the 
mobile $|z\rangle$ orbitals are occupied at all doping levels [see 
Fig.~\ref{ttk_Jse0.02_EJT0}]. In the spectral density depicted in 
Fig.~\ref{Ak_b100_Jse0.02_0h}, one sees signals with low weight for the 
$|x\rangle$ orbitals and a rather broad incoherent peak due to the 
$|z\rangle$ excitations. The broad peak contains dispersionless 
states stemming from $|z\rangle$ electrons next to the few $|x\rangle$ 
electrons. Additionally, a dispersive band is found which comes from 
$|z\rangle$ electrons moving in the AF background, showing a remarkable 
similarity to a one-orbital model \cite{KollerPruell2002a}. 
At the higher temperature $\beta t=30$, more $|x\rangle$ orbitals are 
occupied ($4.53 \pm 0.07 \simeq 38$\% vs. $2.03\pm 0.1$ at $\beta t=100$), 
and their weight in the spectral density increases
[Fig.~\ref{Ak_b30_Jse0.02_0h}]. The dispersionless broad feature becomes 
stronger, as the motion of $|z\rangle$ electrons is hindered by $|x\rangle$ 
electrons. Both effects are further enhanced when spin order weakens at
very high temperature $\beta t=10$.

A chain doped with three holes $x=0.25$ has FM order of
core spins [see Fig.~\ref{ssk_Jse0.02_EJT0}] and practically only mobile
$|z\rangle$ electrons [Fig.~\ref{ttk_Jse0.02_EJT0}], due to the
dominating DE mechanism which favors both FM order and ferro orbital
polarization. The spectral density at low temperature $\beta t=100$ is 
shown in Fig.~\ref{Ak_b100_Jse0.02_3h} and one sees a coherent band with 
almost unrenormalized dispersion $\sim 4t$ due to $e_g$ electrons moving 
in $|z\rangle$ orbitals, and only a weak signal above the Fermi energy 
for the immobile excitations in $|x\rangle$ orbitals. For $\beta t=30$ 
[see Fig.~\ref{Ak_b30_Jse0.02_3h}], the peaks of the $|z\rangle$-band are 
broadened, the bandwidth is slightly reduced and a small spectral weight 
is induced in $|x\rangle$ states below the Fermi energy ($0.12\pm0.02$ 
for the filling by 9 electrons). Again, these effects become more
pronounced for the high temperatures $\beta t=10$, see
Fig.~\ref{Ak_b10_Jse0.02_3h}, where $1.18\pm 0.06$ electrons occupy
$|x\rangle$ orbitals. The broadening of the peaks and the reduction of 
the bandwidth result from core spin fluctuations weakening FM 
correlations, while the additional (incoherent) signals besides the 
tight-binding band are caused by the $|x\rangle$ electrons. The 
reduction of FM correlations and the increasing electron density within 
$|x\rangle$ orbitals occur simultaneously with rising temperature. 

\begin{figure}[htb]
  \centering
  \subfigure{\includegraphics[width=0.33\textwidth]{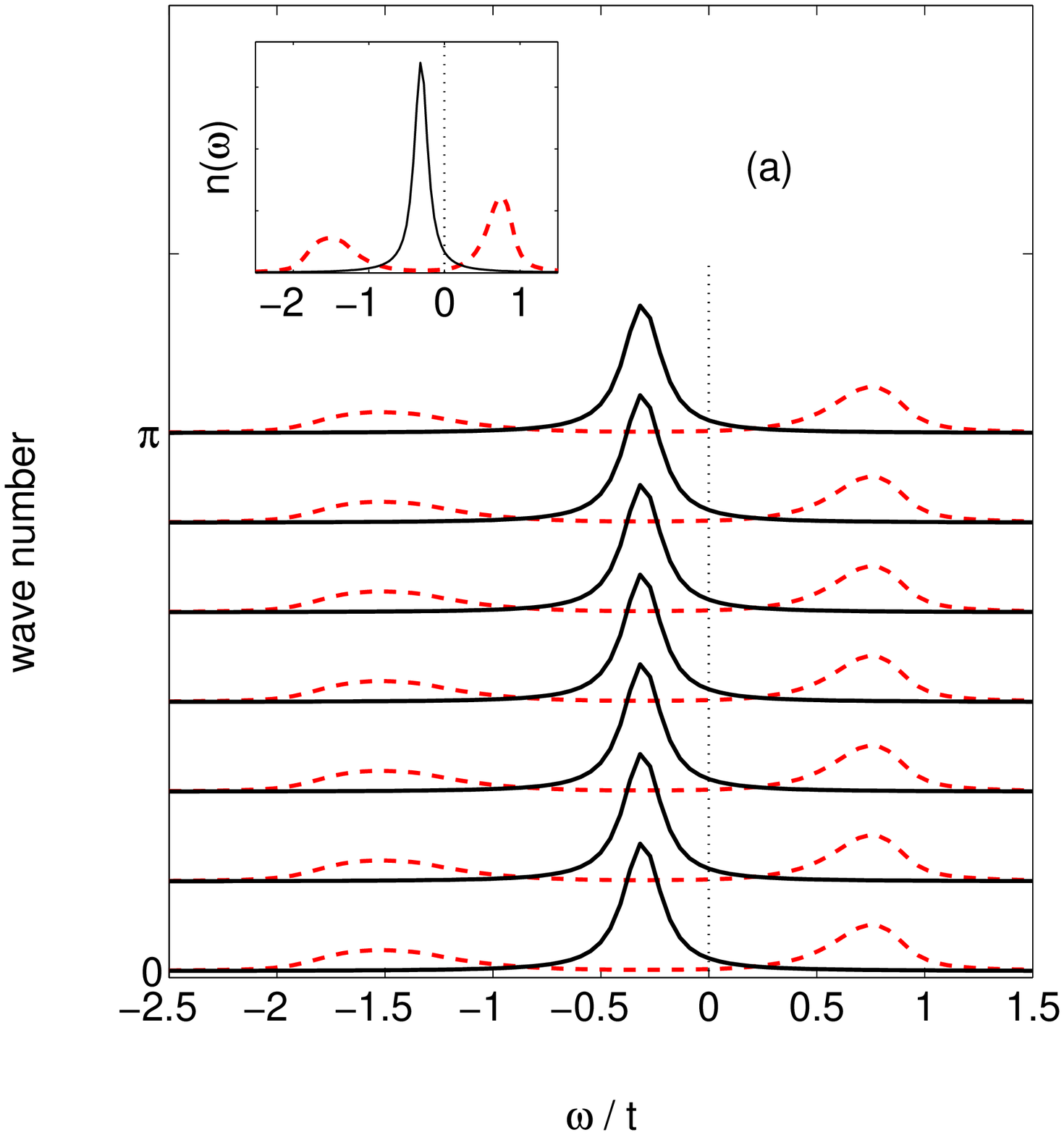}\label{Ak_b100_EJT01_0h}}
  \subfigure{\includegraphics[width=0.322\textwidth]{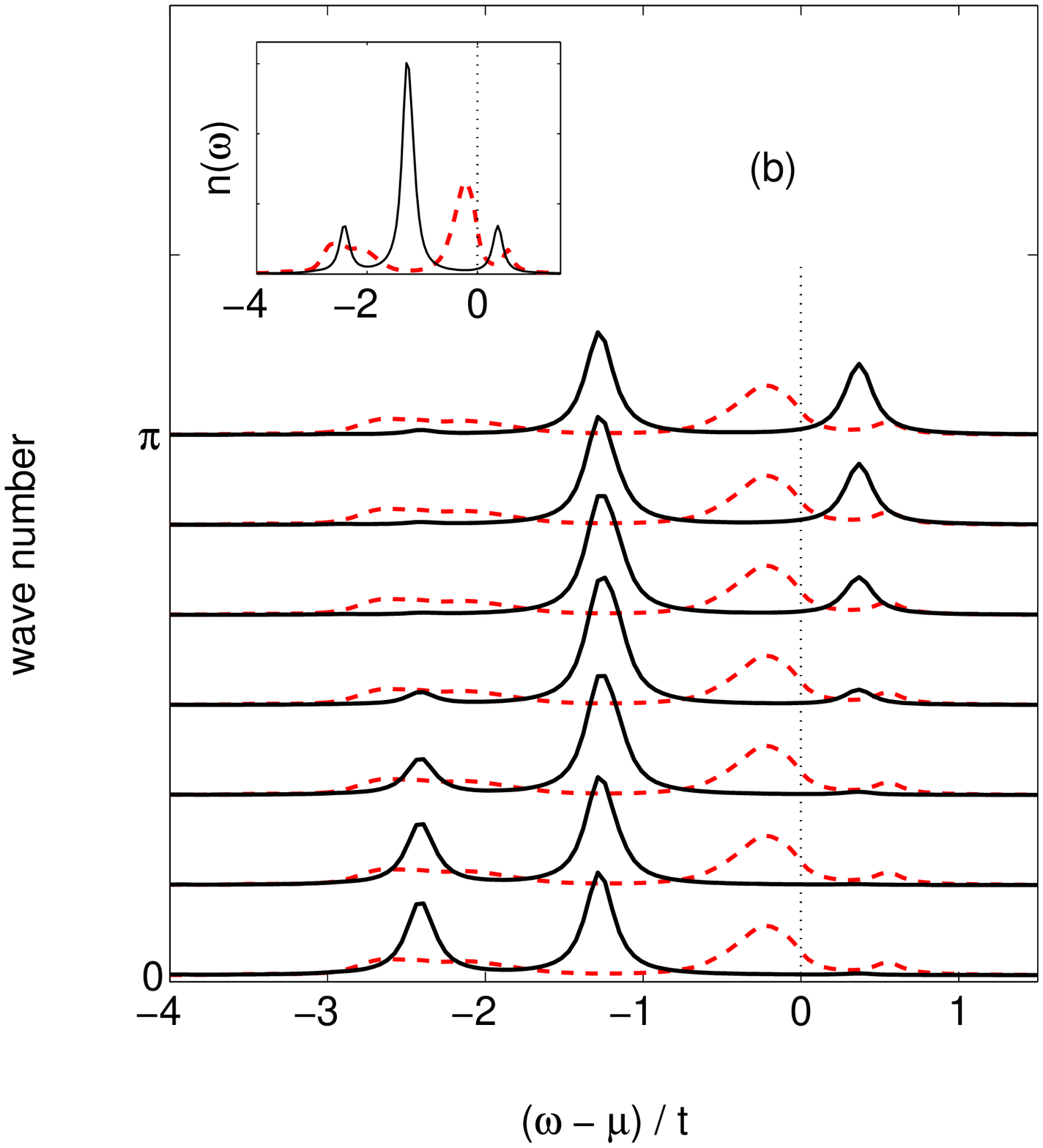}\label{Ak_b100_EJT01_1h}}
  \subfigure{\includegraphics[width=0.325\textwidth]{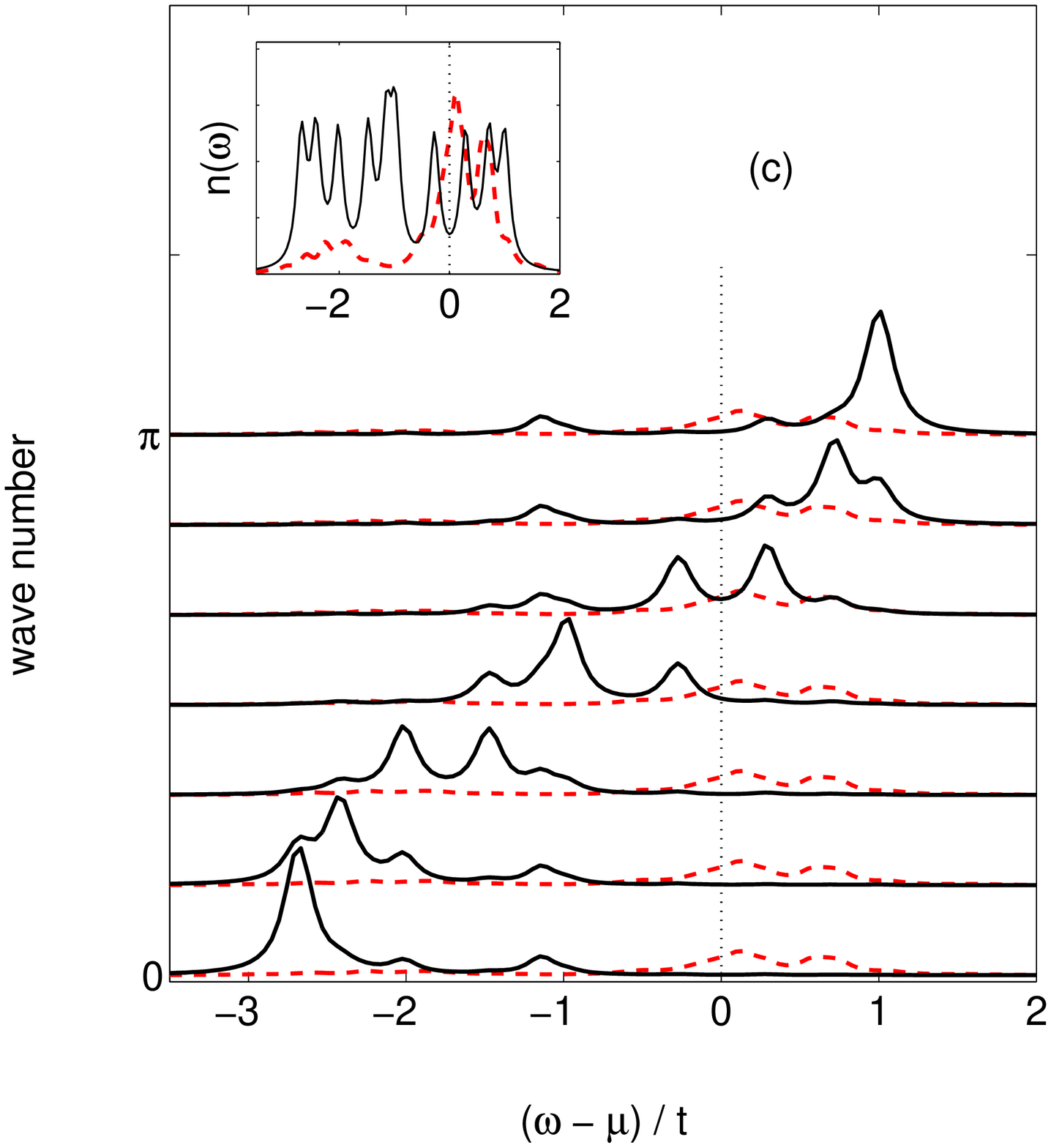}\label{Ak_b100_EJT01_3h}}
  \caption{
  Spectral functions $A(k,\omega)$ for an $N=12$ orbital chain as in 
  Fig. \ref{Ak_Jse0.02}, but with JT potential $E_\textnormal{JT}=0.1t$,
  as obtained for: (a) undoped chain ($x=0$), and chain doped with 
  (b) one hole $x\simeq 0.083$, and (c) three holes $x=0.25$.
  }
\end{figure}

Yet another situation arises (at $J'=0.02t$) in presence of a small 
alternating JT potential $E_{JT}=0.1t$. It favors alternating orbital 
pattern [Fig.~\ref{ttk_Jse0.02_EJT0.1}], which in turn destroys the 
core spin AF order in the undoped chain 
[Fig.~\ref{ssk_Jse0.02_EJT0.1}]. Upon doping, the initially weak FM 
correlations increase, $|x\rangle$ orbitals are depleted and the number 
of $|z\rangle$ electrons even increases slightly to $\sim 7$ for 2 
and 3 holes. For the undoped chain there are just three peaks in the 
spectral function $A(k,\omega)$ [Fig. \ref{Ak_b100_EJT01_0h}]: a 
central peak from the $|z\rangle$ electrons trapped between two occupied 
$|x\rangle$ orbitals, and two side peaks due to the $|x\rangle$ electrons. 
Very similar results (not shown) are obtained for $J'=0$ and 
$E_\textnormal{JT}=0$, where the orbitals also alternate, see above.

Doping with one hole removes an $|x\rangle$ electron and a three-site 
well is created, leading to three $|z\rangle$ signals in the spectral 
density depicted in Fig.~\ref{Ak_b100_EJT01_1h}. The spectral function
$A(k,\omega)$ for doping with three holes [Fig. \ref{Ak_b100_EJT01_3h}] 
shows new features on top of the tight-binding-like band at 
$E_\textnormal{JT}=0$ (Fig.~\ref{Ak_b100_Jse0.02_3h}), that arise from 
the two remaining $|x\rangle$ electrons and from the alternating on-site 
potential seen by $|z\rangle$ electrons. Note that the FM is weaker 
at this doping level than when the JT potential is absent, because DE 
is hindered. From doping $x\geq 1/3$ onward, the chain is fully 
polarized and FM order is enhanced, see Figs. \ref{ssk_Jse0.02_EJT0.1} 
and \ref{ttk_Jse0.02_EJT0.1}. A strong JT potential 
$E_\textnormal{JT}\sim 0.5t$ delays a crossover to the metallic FM 
phase to very high doping $x>0.4$.

Finally, we determined the DE constant for the present 1D chain,
$J_{\rm DE}=\langle H_t\rangle/(2zS^2)$, where $z=2$ and $S=(4-x)/2$
\cite{Ole02}. While the DE increases with $x$ and is simply frustrated 
by the AF SE between core spins ($J_{\rm DE}$ does not depend on $J'$),
increasing JT potential reduces $J_{\rm DE}$ in the insulating regime.
For instance, at $x=0.25$ one finds $J_{\rm DE}\simeq 0.030$, 0.027 and
$0.024t$ for $E_{\rm JT}=0$, 0.1 and $0.25t$. However, at doping  
$x=0.42$ the FM phase is already metallic for the entire range of
$0<E_{\rm JT}<0.25t$, with $J_{\rm DE}\simeq 0.046t$. 

\section{Summary and conclusions}
Even in one dimension, the orbital $t$-$J$ model shows an interesting
competition of FM and AF SE terms. At low temperatures and for the 
undoped chain, a relatively small AF SE $J'$ between the $t_{2g}$ core 
spins already suffices for a transition from FM spin order and alternating 
orbital order ($J'=0$) to AF spin order and ferro orbital polarization 
($J'=0.02 t$). These different situations, both encountered qualitatively 
in LaMnO$_3$, result in quite distinct one-particle spectra. Because of 
the close interplay of orbital- and spin-degrees of freedom, the change 
between these two scenarios can also be triggered by an alternating JT 
potential ($E_{\rm JT}=0.1t$). At higher temperatures, the competing 
interactions lead to further changes in the spectra as both the spin 
and the orbital order are destroyed.

For the doped chain ($x = 0.25$) without JT potential, the kinetic 
energy favors FM order and orbital polarization. This metallic FM
phase may be destroyed either by finite JT potential, or by increasing 
temperature, when both the FM order and the orbital polarization 
are reduced, leading in both cases to incoherent spectra with reduced 
bandwidth. This behavior indicates that an insulating phase may be 
induced either by increasing temperature, or by enhancing the JT 
distortions by chemical substitution. Summarizing, the present study
highlights: ($i$) a complex interplay between spin, orbital and change
degrees of freedom, and ($ii$) the importance of the coupling to the lattice
--- all these all these aspects decide about spin and orbital correlations in
doped manganites and have to be included in their realistic microscopic
models\cite{Weisse_04}. 

\begin{acknowledgement}
This work has been supported by the Austrian Science Fund (FWF), 
Project No.~P15834-PHY, and by the Polish State Committee 
of Scientific Research (KBN), Project No.~1 P03B 068 26.
\end{acknowledgement}

\bibliographystyle{unsrt}
%% \bibliography{manganites}

\end{document}